# Electron Beam Radiolysis-Assisted Growth of Rutile $TiO_2$ Thin Films

*Silu Guo[1#], Nitin Sathish Kumar[1#], Sreejith Nair[1#], Supriya Ghosh[1], Bharat Jalan[1] and K. Andre Mkhoyan[1]*

[1]Chemical Engineering and Materials Science, University of Minnesota, Twin Cities, Minneapolis, MN 55455, USA

[#] Equal contribution

**Abstract**

A new approach for growing crystalline thin films is developed that takes advantage of electron beam radiolysis being a constructive force to rearrange atoms into a crystalline structure. It is demonstrated that by irradiating the surface of a $TiO_2$ film by an electron beam supplied by a reflection high energy electron diffraction (RHEED) gun inside the MBE chamber during growth, a crystalline film can be grown at much lower substrate temperatures, where deposited films typically appear amorphous. Here, rutile $TiO_2$ films were grown using hybrid molecular beam epitaxy (MBE) allowing atomic level control of growth as well as an observation of radiolysis-driven crystallization. Analysis was carried out using a combination of SEM and atomic-resolution STEM imaging. It is also shown that by tuning the temperature of the substrate and the dose of the electron beam, the degree of crystallinity of the film can be controlled.

## Introduction

Epitaxial growth of crystalline thin films requires energy to promote adatom movement on the surface during film growth[1,2]. The most common way of providing needed energy is by adjusting the substrate temperature to a particular value for the material that is intended to grow[3,4]. While elevating the substrate temperature is typically the solution, an excessive thermal energy can be counterproductive resulting in a dramatic reduction of atomic adhesion[2]. Other known approaches reported in literature that promote crystalline film growth include incorporation of an in-plane electric field in the substrate during growth[5], ion-beam bombardment[6], and acoustic activation of the substrate[7]. When needed energy is provided, it induces crystallinity in the growing film allowing atomic motions and sufficient surface diffusivity[2]. If the energy is not sufficient, or if the substrate temperature is below the required value, the limited diffusion of adatoms often result in the growth of an amorphous film instead[2,8]. The effect of temperature on crystallization of the film is studied in a variety of materials including transition metal perovskite oxides like $SrTiO_3$[9], or binary oxides like $VO_2$[10] or $TiO_2$[11].

It should be noted that reports in the literature also indicate that changes in the atomic structure of the growing films can be introduced by bombarding the region with electron beam, resulting in improvements in crystal structure[12], surface coverage[13] and smoothness of the growing films[14,15]. The effects of electron beam irradiation can be divided into two primary categories: knock-on effects where atoms are displaced due to directs electron-atom collisions, and radiolysis where atoms are moved to due relaxation of local structure and consumption of exciton energy[16,17]. While knock-on effects are almost always destructive for atomic structure of the material, the effects of radiolysis could be destructive as well as constructive. Atomic restructuring via electron beam radiolysis has been observed in a variety of materials[16] including silicates[18,19], halides[20,21], and zeolites[22,23,24]. Radiolysis provides rearrangement of atoms under an electron beam, promoting transitions between crystalline and amorphous phases. A recent scanning transmission electron microscopy (STEM)-based study showed that radiolysis in $TiO_2$, which is a material that generates stable high-energy excitons in the presence of an electron beam[25,26,27], leads to atomic movements that allow healing of cracks in the structure[25].

In the study presented here, we show that by incorporating electron beam-induced radiolysis, an alternative thin film growth approach can be established where crystalline film can be grown at considerably lower substrate temperatures. This is demonstrated for a $TiO_2$ thin film grown on a rutile $TiO_2$ substrate using molecular beam epitaxy (MBE). MBE provides an atomically precise epitaxial film growth that allows leveraging the effects of electron-beam driven radiolysis which is also an atom-scale process. This specific material system was chosen because $TiO_2$ shows both, a temperature dependent crystallization as well as effects of electron beam radiolysis[25].

**Results and Discussion**

$TiO_2$ thin films were grown epitaxially on a $TiO_2$ (001) substrate using hybrid MBE. Films were grown at different substrate temperatures in the presence of an electron beam supplied by a reflection high energy electron diffraction (RHEED) gun, incident on the film surface at a grazing incidence angle of 1-2°, as depicted schematically in Fig. 1(a). Ti was supplied using a titanium tetraisopropoxide (TTIP) metal-organic precursor while oxygen was supplied in the form of oxygen plasma. The electron beam was incident on the film surface for the entire duration of growth to promote electron beam radiolysis and subsequent atomic restructuring. The schematic depicted in Fig. 1(b) shows a grown film exhibiting localized structural differences due to radiolysis.

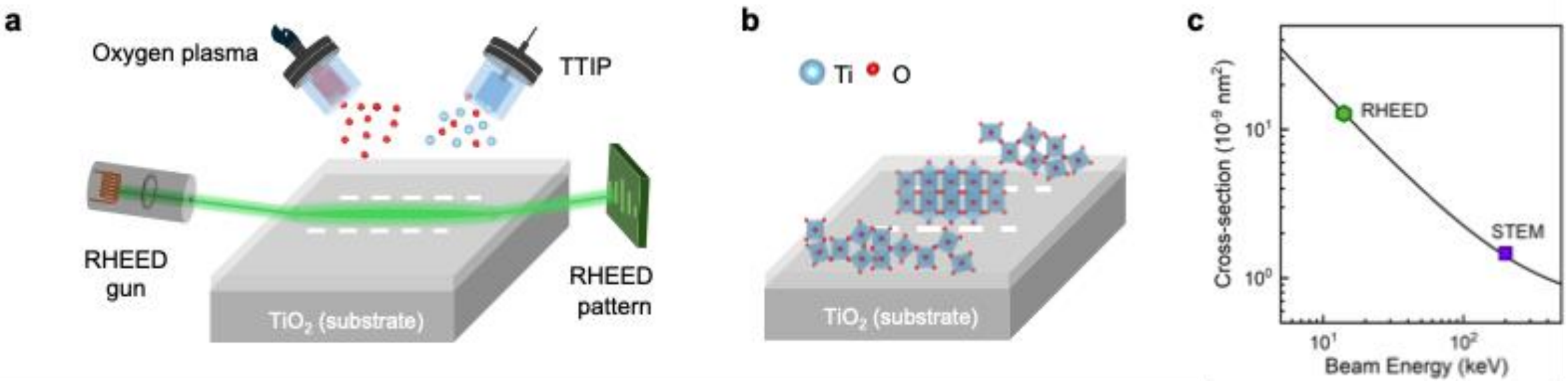

**Fig. 1. Electron beam-assisted film growth. (a)** Schematic of film growth in the presence of an electron beam supplied by the RHEED gun. **(b)** Schematic description of localized crystallinity in the film due to electron beam radiolysis during growth. **(c)** Radiolysis cross-section as function of electron beam energy.

The radiolysis cross-section ($\sigma_r$) has an increasing trend with a decrease in incident electron beam energy, $E_0$[16]:

$$\sigma_r(E_0) = 8\pi a_0^2 \times \left(Z \frac{U_R}{m_0c^2}\right)\left(\frac{U_R}{E_{th}\beta^2}\right) \times \zeta \,, \tag{1}$$

where $a_0$ is the Bohr radius, $Z$ is the average atomic number of a moving unit, $U_R$ is Rydberg energy, $E_{th}$ is the threshold energy required to induce atomic movement, and $\zeta$ is the efficiency of radiolysis. The factor $\beta = \frac{v}{c} = \sqrt{1-(1+\frac{E_0}{m_0c^2})^{-2}}$, where the inverse relation to beam energy, $E_0$, is present. Fig. 1(c) shows this trend indicating that the RHEED beam with typical energy of $E_0$ = 10-20 keV has about 10x higher radiolysis cross-section than a typical STEM beam with beam energy of $E_0$ = 200-300 keV.

A set of five $TiO_2$ films were grown at substrate temperatures of 100 °C, 130 °C, 140 °C, 150 °C and 300 °C, all in the presence of $E_0$ = 14 keV electron beam supplied by the RHEED gun. The surfaces of as-grown films were imaged using a Scanning Electron Microscope (SEM) in secondary electron (SE) mode to look for any features of radiolysis induced restructuring. Fig. 2(a) shows the SEM images of these as-grown film surfaces for the films grown at 100 °C, 130 °C, 140 °C, and 150 °C, and Supplementary Fig. S2 (c) shows the SEM image of the 300 °C grown film surface when viewed from the top. Corresponding RHEED patterns obtained before and after film growth are shown in Supplementary Fig. S1. The films grown at the low and high ends of the growth temperatures, i.e., 100 °C and 300 °C, show no visible feature in these SEM images (for SEM image from 300 °C grown sample see Supplementary Fig. S2). When analyzing these films in cross-sections using HAADF-STEM imaging, it was observed that the film grown at 100 °C showed an amorphous atomic structure throughout, and the film grown at 300 °C showed a completely crystalline structure (see Supplementary Fig. S2). On the other hand, the SEM images of the films grown at 130 °C, 140 °C, and 150 °C (Fig. 2 (a)) show a vertical rectangular strip-like feature at the location of electron beam exposure. The feature passes across the entire length of the film, has a width of about ~ 200 μm, and exhibits a brighter image intensity compared to the surrounding regions of the film. The position and orientation of this feature is marked using vertical white dashed lines on all images shown in Fig. 2(a).

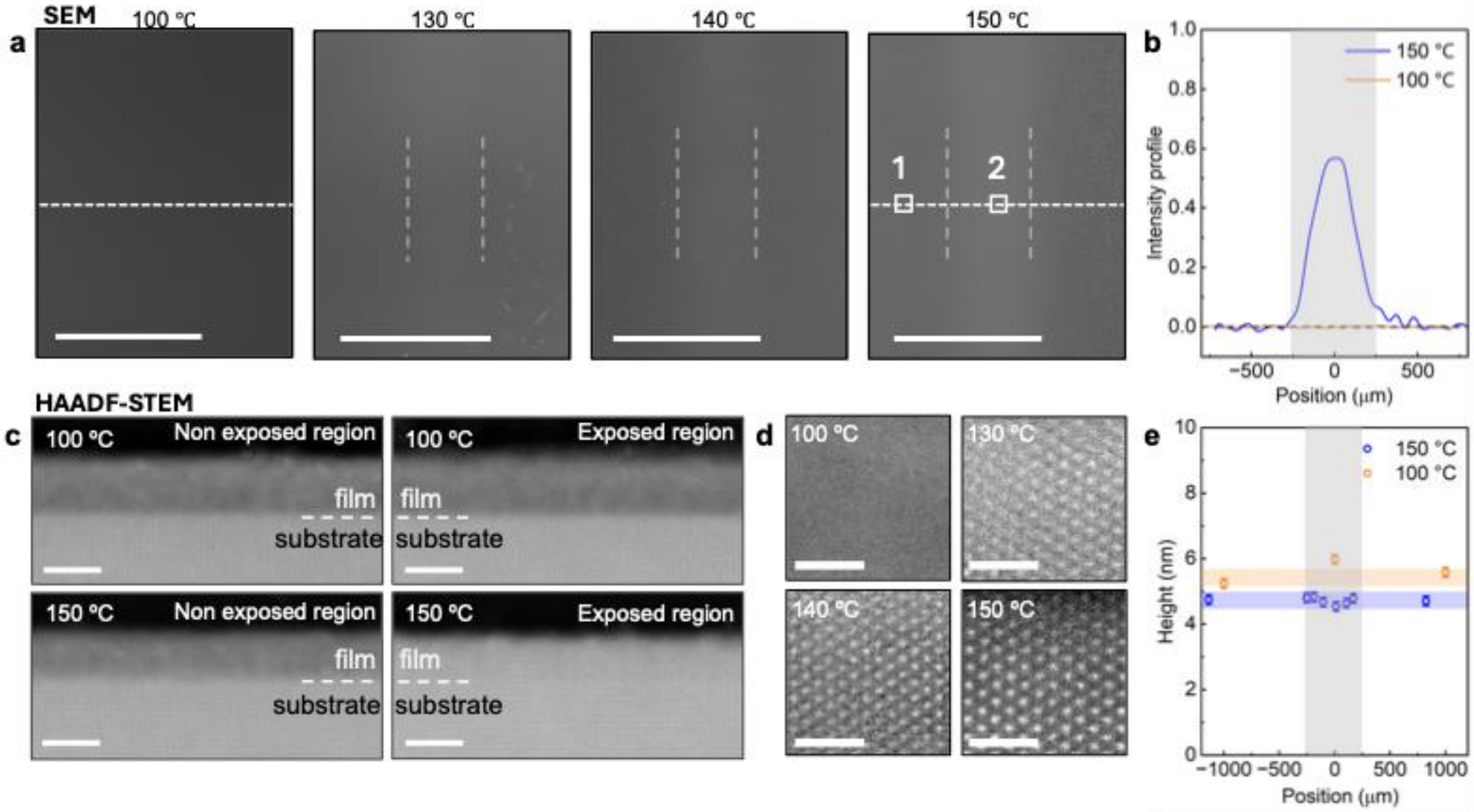

**Fig. 2. Comparison of film crystallinity and height between electron beam exposed and non-exposed regions. (a)** A set of SEM images from as-grown film surfaces grown at 100 °C, 130 °C, 140 °C, and 150 °C substrate temperatures. Scale bars are 500 μm. All films were exposed to a strip of the RHEED beam during film growth illustrated in Fig. 1(a). The vertical white dashed lines indicate the region of electron beam incidence. **(b)** Intensity line-scan profiles of SEM images of films grown at 100 °C and 150 °C taken across electron beam exposed region. (**c)** HAADF-STEM images of film cross-sections for the films grown at 100 °C and 150 °C from electron beam-exposed and non-exposed regions. A non-exposed region is selected from a location far from the region of electron beam incidence, marked by box 1, and an exposed region is selected from location, marked by box 2 in panel (a). Scale bars are 5 nm. **(d)** High resolution HAADF-STEM images of film cross-sections grown at 100 °C, 130 °C, 140 °C, and 150 °C substrate temperatures. Scale bars are 1 nm. **(e)** Film heights extracted from cross-section STEM images at different positions with respect to the center of the RHEED beam incidence region for films grown at 100 °C and 150 °C. The blue and orange bands have a width of one unit cell of $TiO_2$, 4.6 Å.

The contrast of this feature increases with an increase in the growth temperature; the film grown at 130 °C showing the faintest contrast, and the film grown at 150 °C showing a very evident one. Further, for each temperature, the feature is such that its intensity is the highest at the center of the beam incidence region and gradually fades away as we move away from the beam exposed region towards the non-exposed regions, corresponding to the intensity profile of the RHEED beam (Fig. 2. (b)) obtained after acquiring an intensity line-scan along the horizontal length of the SEM image (shown by horizontal dashed lines in the SEM images of the 100 °C and the 150 °C films in Fig.

2. (a)). This feature in the SEM image can be attributed to either crystallinity differences, i.e., a contrast arising because of a difference in the crystallinity of the grown film in different regions, or a surface topography, i.e., contrast arising due to variations in the film height or roughness, between the beam exposed and non-exposed areas, or a combination of two[28].

To differentiate the effects of temperature and radiolysis on the structure of the films, cross-sectional samples from films grown at 100 °C (with no visible features in SEM image, Fig. 2(a)) and 150 °C (with the brightest contrast in SEM image, Fig. 2(a)) were prepared and studied using HAADF-STEM imaging. As can be seen in Figs. 2(c) and (d), in 100 °C grown film, both the beam exposed and non-exposed regions are structurally amorphous. However, in 150 °C grown film, while non-exposed region is structurally amorphous, the RHEED beam exposed area is crystalline. Additional HAADF-STEM imaging from 130 °C and 140 °C grown sample showed that the films grown at these temperatures also have crystalline order in them (Figs. 2(d)). The analysis of the local film height (or thickness) showed that variation is very low, less then 0.7 nm, which is at the level of 1–2-unit cells of rutile $TiO_2$ (Fig. 2(e)). Similar results were observed for local surface roughness (see Supplementary Fig. S3). It should be noted that all the $TiO_2$ films grown at different substrate temperatures discussed above showed similarly low local film height variation as well as surface roughness (see Supplementary Fig. S3). These observations suggest that the main contributor in the SEM image contrast is the crystallinity of the local area, rather than the surface topography.

The results from HAADF-STEM imaging shows that, while the electron beam exposed regions of the films grown at 130 °C, 140 °C and 150 °C have crystalline order in them, and the degree of this order is different in different films (Fig. 2(d)). For quantification of local crystallinity of the $TiO_2$ films, we introduce a quantity referred here as degree of crystallinity:

$$S = \frac{\sum_{i=1}^{n} I_B^f}{I_Z^f} \frac{I_Z^s}{\sum_{i=1}^{n} I_B^s} \qquad (2)$$

which evaluates the average intensity of $n$ equidistant Bragg spots in the Fast Fourier Transform (FFT) of the HAADF-STEM image of the film, $\frac{1}{n}\sum_{i=1}^{n} I_B^f$, with respect to that in fully crystalline substrate, $\frac{1}{n}\sum_{i=1}^{n} I_B^s$. To minimize the effect of thickness differences between the film and substrate, and between samples with different thicknesses, the intensity of the average Bragg spot is normalized to the intensity of "Zero" spot of the FFT, $I_Z^{f/s}$.

Such quantification of crystallinity of $TiO_2$ films grown at 150 °C is shown in Fig. 3. The yellow and blue boxes on the left panels for both Fig. 3(a) and (b), show atomic-resolution HAADF-STEM images of regions on the film and substrate respectively. Calculated FFTs from those images are shown in the insets of the top-right and bottom right panels of Fig. 3(a) and (b). Six Bragg spots corresponding to the first-order frequencies were identified in the FFTs of the film and the substrate. They are marked by red circles in each FFT as shown in Fig. 3(a) and (b). The intensity of each Bragg spot was obtained by integrating the intensity within these disks. To isolate the periodic order, a local background subtraction was performed. The background was estimated at the same distance of the Bragg spots from the central spot, but at angular positions midway between adjacent Bragg spots to get the background intensity under equivalent frequency while avoiding peak overlap. The "Zero" frequency component, $I_Z^{f/s}$, is obtained by integrating the FFT intensities within a disk of the same size centered at the origin. The disk size was chosen to correspond to a characteristic real-space length of 1 nm, reflecting the minimum length of structural variations in the analyzed images, presented as patches of the same length scale.

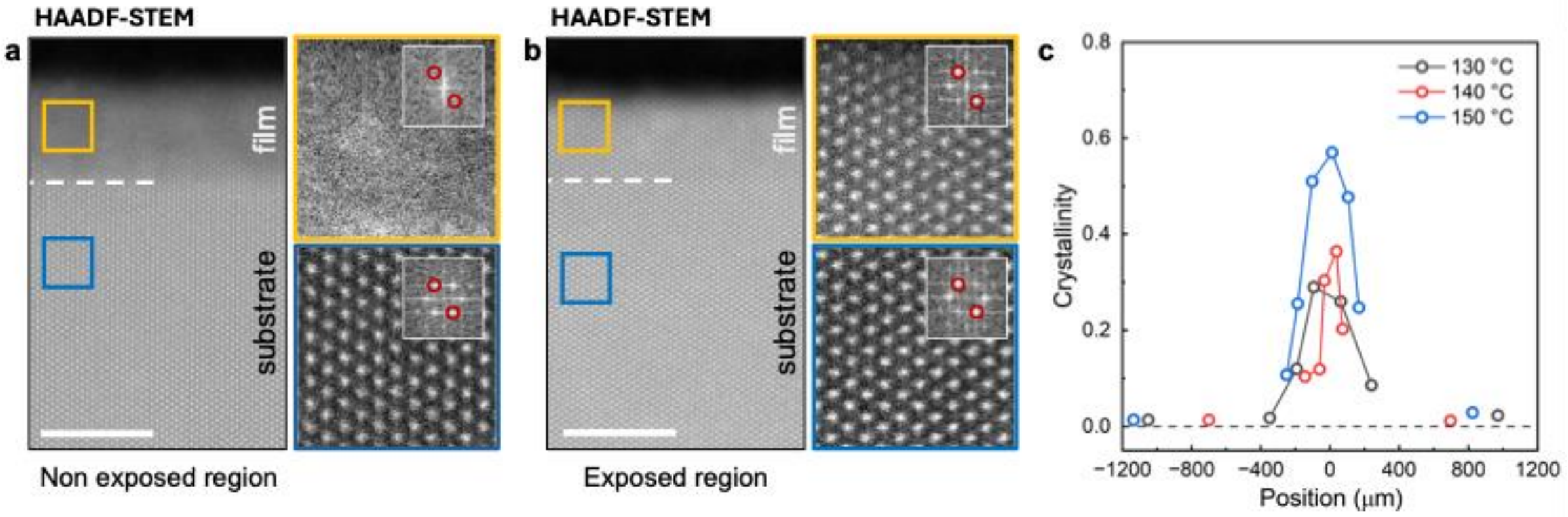

**Fig. 3. Degree of Crystallinity in the film and its variation. (a)** (left) A cross-section HAADF-STEM image of the film grown at 150 °C, from the region in box 1 in Fig. 2(a). The top-right panel with a yellow boundary is an image of the region on the film in the left panel marked with a box of the same color, with the inset showing the corresponding FFT. The bottom-right panel is an image with a blue boundary that marks a similar region on the substrate, with the inset showing the corresponding FFT. **(b)** A cross-section HAADF-STEM image of the film grown at 150 °C,

from box 2 in Fig. 2(a). Scale bars for (a) and (b) are 5 nm. **(c)** The degree of crystallinity of the film from different locations relative to the RHEED beam incidence region for films grown at 130 °C, 140 °C and 150 °C.

The results of quantification of crystallinity for the films that show a feature in SEM images due to irradiation of the RHEED beam, i.e., those grown at 130 °C, 140 °C, and 150 °C, are presented in Fig. 3(c). For each case, the crystallinity increases as we move to the center of the RHEED beam exposed region, i.e., to a region of higher electron dose. Also, the degree of crystallinity in the RHEED beam exposed area increases with substrate temperature, with the 150 °C grown film having the highest crystallinity at the center, and 130 °C grown film having the least. These observations corroborate with the RHEED patterns measured during the film growth (see Supplementary Fig. S1). The "rise and fall" for the degree of crystallinity in the beam exposed area is due to focused RHEED beam intensity distribution, consistent with SEM observations (Fig. 2(b)). The crystallinity peaks at the location where the electron dose is the highest, i.e., at the center of the beam, and gradually drops as we move to the non-exposed regions.

The results presented above showing that for films grown at temperatures in the range 130 °C to 150 °C only electron beam exposed areas are crystalline and, within that area the degree of crystallinity follows the RHEED beam profile, directly confirming that indeed the radiolysis is the driving force behind crystallization of the beam exposed regions of the film. However, since $TiO_2$ films grown at higher temperatures crystalize more effectively under electron beam exposure, it can be concluded that the radiolysis provides needed energy for crystallization when it is not available at low substrate temperature. For $TiO_2$ films grown at 100 °C the radiolysis from 14 keV RHEED beam appears to be insufficient, as a result even beam exposed area of the film is not crystalline.

As discussed earlier, the efficiency of radiolysis can be increased by lowering electron beam energy or by increasing the dose (see Fig. 1(c) and Eq. (1)). To demonstrate that the crystallinity improves with an increase in the electron dose, a STEM probe scanning experiment at room temperature was carried out on cross-section samples prepared from the non-exposed region on the $TiO_2$ film where the film is amorphous. The results are presented in Fig. 4. As can be seen from Fig. 4(a) where HAADF STEM image of a film cross-section is shown (obtained from box 1 region in Fig. 2(a)), the film is indeed amorphous before STEM beam exposure. The same region of the film after continuous scanning with a STEM probe for an hour appears brighter and with crystalline atomic structure. Atomic resolution HAADF-STEM images obtained at different doses (Fig. 4(b)) shows this radiolysis-driven amorphous-to-crystalline transformation of the $TiO_2$ film. It also shows that the degree of crystallinity of the film increases with increase of the electron dose. The analysis of degree of crystallinity of STEM beam exposed region as a function of electron dose is presented in Fig. 4(c) showing linear dependence with electron dose. These results further confirm that such radiolytic amorphous-to-crystal transformation can be improved by increasing the dose of the electron beam.

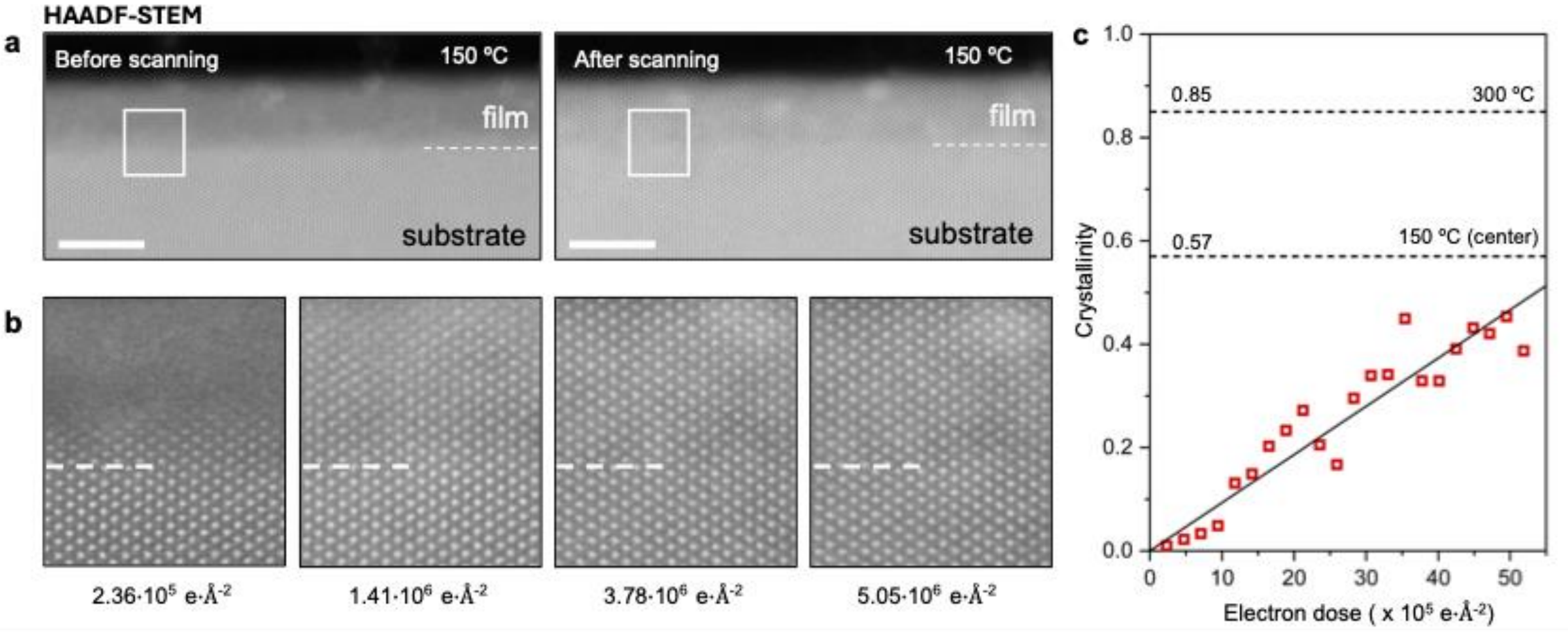

**Fig. 4. Radiolysis-driven crystallization of amorphous $TiO_2$ film under STEM beam exposure. (a)** HAADF-STEM images of the $TiO_2$ film cross-section prepared from a non-exposed region of the film, grown at 150 °C, obtained before (left panel) and after (right panel) continuous STEM beam scanning. Scale bars are 5 nm. **(b)** Evolution of the film crystallinity with increasing electron dose. Images are from a region marked by the white box in (a). The dashed lines indicate film-substrate interface. **(c)** Profile of the degree of crystallinity with respect to the electron dose exhibiting a linear trend.

In conclusion, the results presented here show that when the electron beam irradiates the surface of a growing $TiO_2$ film inside the MBE chamber, radiolysis effects are strong enough to transform otherwise amorphous $TiO_2$ into a crystalline film. It was shown that when $TiO_2$ film is grown at lower temperature, ranging from 100 °C, 130 °C, 140 °C, 150 °C, the thermal energy provided from substrate is not sufficient to grow crystalline films. However, when electron beam exposure is present, the films will grow crystalline even at these low temperatures (excluding 100 °C case). These results suggest that films, which can sustain electron beam radiolysis, can be grown with a wider range of substrate temperatures. It also suggests that materials with lower melting points can also be considered as a substrate for film growth by complimenting low temperature heating with electron beam exposure. Since radiolysis is the driving force, the quality of the crystallinity of the film can be adjusted by controlling the dose of the electron beam. While the results presented here are from experiments conducted with MBE growth of the films, it should be applicable for all film growth techniques. The efficiency of this radiolysis-assisted approach can be further improved by lowering electron beam energy, since it is inversely proportional to beam energy.

## Methods

### MBE growth of $TiO_2$ films

Epitaxial $TiO_2$ (001) thin films were grown on single crystal rutile $TiO_2$ (001) substrates using hybrid molecular beam epitaxy technique[29,30], while constantly exposing the sample to an electron beam from a RHEED (Staib Instruments, USA) electron gun. The single crystal substrate (Crystec GmbH) was first cleaned sequentially in acetone, methanol and isopropanol to remove any organic residue. The substrate was then loaded into an ultra-high vacuum sample preparation chamber and baked at 200 °C for 2 hrs to eliminate volatile adsorbed species. The substrate was then transferred to a synthesis chamber while maintaining ultra-high vacuum for subsequent thin film deposition. The substrate is heated to the desired growth temperature (100-300 °C in this work). Note the thermocouple is placed adjacent to the heater and not in direct contact with the substrate, so the actual substrate temperature is expected to be lower than the reported substrate temperature. Once at the desired growth temperature, the substrate is exposed to oxygen plasma, which is generated using an inductively coupled radiofrequency plasma system operated at a forward power of 250 W. The oxygen plasma exposure further helps remove any adsorbed hydrocarbon species and improve the surface quality of the substrate. Before initiating thin film growth, the substrate is characterized in-situ using an electron beam from the RHEED gun. A RHEED filament current of 1.5 A and an acceleration voltage of 14 kV is used throughout this work. The beam focus voltage is kept constant throughout and only the grid voltage is adjusted to tweak the beam current for obtaining a bright and sharp diffraction pattern, a standard practice in MBE synthesis. Due to technical limitations in measuring the beam current, we rely on an assumed beam current of 100 nA for all calculations.

### SEM characterization and preparation of cross-section lamella

The as-grown films were attached to SEM stubs mounted onto an SEM specimen holder and loaded into the system for imaging the surface of the as-grown film. The SEM images were captured using a 5 keV electron beam with a beam current of 25 pA. After identification of the RHEED beam feature, their edges were marked by “crosses” that acted as fiducial marks, made via shallow trenching using the Ga ion beam, as the feature itself gets hidden once the film surface is coated with amorphous carbon for cross-section sample preparation. These marked films are then coated with a layer of 50 nm thick amorphous carbon that acts as a protective coat for the film underneath for ion-beam milling. Cross-section samples were prepared from these amorphous carbon coated $TiO_2$ film surfaces using a Ga ion beam in a FEI Helios G4 dual-beam FIB-SEM system. Software built-in patterns were used to trench the surface of the film using an ion beam at 30 kV with the beam current ranging from 7 pA to 9.1 nA. A cross-section lamella of length spanning 10 μm, height of 3-4 μm, and a thickness of 1μm was cut out of the trench and welded onto a TEM half grid using Pt. The lamella was further thinned down so that a thickness of 60 nm or lower is achieved to ensure electron transparency during STEM imaging of the cross-section.

**Surface height and roughness analysis**

Multiple HAADF-STEM images of the film cross-section from regions on the electron transparent cross-section samples were acquired using an aberration-corrected FEI Titan G2 60–300 (S)TEM microscope equipped with a CEOS DCOR probe corrector, a Schottky field emission gun, a monochromator, a super-X energy dispersive X-ray (EDX) spectrometer, and a Gatan Enfinium ER spectrometer. The acquired images were then annotated using Gatan DigitalMicrograph software using annotations with calibrated length markers, after identifying the horizontal plane of atoms that lie at the film-substrate interface. These annotations were made such that they started from this interface and extended vertically to the top of the film surface, with 40 such vertical length markers covering the horizontal length of the film equidistantly. The obtained histogram showed a gaussian fit, with the mean height determined from the peak position of the fit. The analysis was carried out for 4-5 images taken at different regions on each film cross-section, and the mean height of the film at the location the cross-section sample was prepared is calculated. The same analysis was carried out for samples prepared at different positions on the film surface. A surface roughness vs position is obtained by recording the standard deviation ($\sigma$) of the histogram gaussian fit obtained for each cross-section sample. A value of $3\sigma$ was chosen to quantify surface roughness.

**STEM scanning experiments**

Electron-beam-assisted crystal growth experiments were conducted on a FIB cross-sectional sample taken from the amorphous region of the 150 °C-grown film. STEM imaging was performed at an accelerating voltage of 200 kV using a probe current of 90 pA. Images were acquired with a dwell time of 8 μs per pixel and a frame size of 2048 × 2048 pixels, corresponding to a total acquisition time of 40.3 s for each frame. The scanned area is 31 × 31 $nm^2$. The electron dose rate was calculated as the incident electron flux normalized to the scanned area. Under these imaging conditions, the electron dose rate was $5.85 \times 10^4$ e $Å^{-2}$ $s^{-1}$, corresponding to an accumulated dose of $2.36 \times 10^5$ e $Å^{-2}$ per frame.